# Annealing Relaxation of Ultrasmall Gold Nanostructures

Vitaly Chaban[1]

MEMPHYS – Center for Biomembrane Physics, Syddansk Universitet, Odense, 5230, Kingdom of Denmark

**Abstract**. Except serving as an excellent gift on proper occasions, gold finds applications in life sciences, particularly in diagnostics and therapeutics. These applications were made possible by gold nanoparticles, which differ drastically from macroscopic gold. Versatile surface chemistry of gold nanoparticles allows coating with small molecules, polymers, biological recognition molecules. Theoretical investigation of nanoscale gold is not trivial, because of numerous metastable states in these systems. Unlike elsewhere, this work obtains equilibrium structures using annealing simulations within the recently introduced PM7-MD method. Geometries of the ultrasmall gold nanostructures with chalcogen coverage are described at finite temperature, for the first time.

[1] E-mail: vvchaban@gmail.com

TOC Image

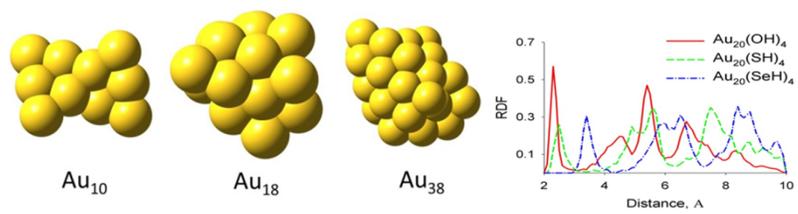

**Research Highlights**

(1) Semiempirical molecular dynamics simulations were carried out.

(2) Annealing was used to obtain relaxed structures for gold nanoclusters.

(3) Gold nanoparticles were described in terms of radial distribution functions.

**Introduction**

Gold, Au, is among those residents of the Periodic Table, which do not require extensive introduction. Its role in human civilization ranges from economy and military history to health science applications.[1-11] The latter direction has recently emerged and employs nanoscale Au for therapeutics and diagnostics due to a drastic alteration of physical chemical properties, spectroscopic, and reactivity behavior at the nanoscale.[12, 13] The present work discusses ultrasmall nanostructures of gold and their functionalization using mathematical modeling at the electronic structure level of accuracy and precision. We will refer to these nanoparticles as to nanoclusters to underline their subnanometer dimensions.

Surface plasmon resonance (SPR) properties foster interest to gold nanostructures.[13] SPR is an optical phenomenon arising from the interaction between an electromagnetic wave and the conduction band electrons of the metallic systems. The conduction electrons of Au under the irradiation of light are driven by an electric field to collectively oscillate at a resonant frequency relative to the ionic lattice. At this resonant frequency, the incident light can be absorbed. Some of these photons will be released with the original frequency in all directions (light scattering process). Furthermore, some of these photons are converted into phonons or vibrations of the ionic lattice (light absorption process). The surface plasmon resonance peak in the case of a gold nanostructure consists of the scattering and absorption components. Size, geometry, and genuine structure of Au nanoclusters determine intersection and ratio of the scattering the adsorption components. Per classical example, in Au nanospheres ca. 50 nm in diameter, the SPR peak is positioned at 520 nm.[13] This peak provides a fancy ruby red color of conventional gold colloid solutions. To recapitulate, size and shape of Au nanostructures play a major role in their optical properties. It is of fundamental importance in the field to be able to predict, control and characterize these structures using both experimental manipulation techniques and computational tools in chemistry.

While the experimental techniques allow to control Au nanostructure size down to one nanometer (which is encouraging though not enough), theoretical predictions of equilibrium geometries are far from being trivial.[12] First, this is due to 79 electrons per atom implying a rich conformational flexibility of even few-atom clusters. Second, conventional geometry relaxation algorithms, such as steepest descent and conjugate gradient, operate poorly for significantly large structures. Third, the relaxed geometry in the absence of kinetic energy (0 K) is expected to be close to the room-temperature geometry in the case of solids, although it is not guaranteed a priori. In this work, equilibrium structures of the pristine and functionalized Au nanoclusters were obtained through annealing simulations using the PM7-MD method,[14] as described in the methodology section.

Nearly all theoretical data on nanoparticles and nanoclusters, which are available in literature thus far, were obtained using density functional theory (DFT) calculations coupled with geometry relaxation algorithms (i.e. consequent wave function optimizations and quasi-Newton structure alterations).[15-17] That is, these structures correspond to the temperature of absolute zero and principally exclude any possible thermal (entropic) effects. This may be a primary reason why not all of the theoretically obtained symmetries were confirmed by the X-ray measurements as outlined by Jiang.[12] Jalkanen and coworkers employed plane-wave periodic DFT to investigate adsorption of tiny gold and silver nanoclusters, such as $X_2$, $X_6$, $X_{13}$, $X \equiv Au, Ag$, on the (0001) graphite surface.[18] The adsorbate atoms favor atop and bond sites, but avoid hollow sites. It was suggested that large structural changes in octahedral $X_6$ and icosahedral $X_{13}$ clusters take place upon adsorption. Interestingly, dispersion forces between metal and carbon atoms were concluded to be important in this case. Tlahuice-Flores and coworkers explored the structure of $Au_{130}$-thiolate and $A_{130}$-dithiolate clusters using rapid electron diffraction in scanning/transmission electron microscopy and DFT-optimized structures.[16] Time-dependent DFT was used to record distinctive optical absorption spectrum, and rationalize the specific stability underlying the selective formation of the $Au_{130}$ cluster in high yield. Similarities and differences with respect to other experimentally known gold containing nanoclusters, $Au_{102}$ and $Au_{144}/Au_{146}$, were outlined.[16] There are some theoretical works, which are devoted to the nanoclusters of other noble elements, such as Ag, Pt.[15, 19]

This report considers efficiency of annealing simulations in the case of relatively small $Au_{10}$, $Au_{18}$, $Au_{38}$ nanoclusters (Figure 1) and three $Au_{20}$ nanoclusters protected by four chalcogen containing groups, –OH, -SH, and –SeH, each. The implemented protection is not complete, since the goal of the work is to characterize structure, while maintaining relatively simple atomistic models.

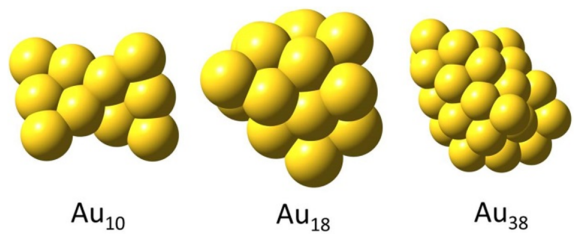

Figure 1. Atomistic configurations of the selected ultrasmall gold nanoclusters, $Au_{10}$, $Au_{18}$, $Au_{38}$, after simulation of annealing (see Table 1 for simulation details). Unlike $Au_{18}$ and $Au_{38}$ nanoclusters, $Au_{10}$ maintains an almost planar structure. This structure remains essentially unaltered during high-temperature spontaneous dynamics of the system.

Table 1. The list of the simulated systems. I provide a total number of electrons per system for a straightforward comparison of computational load with alternative electronic structure studies. Note, that PM7 uses effective-core potentials for all elements except hydrogen and helium. Each gold atom contains 11 explicit electrons. The annealing starting temperature was set significantly smaller for the case of functionalized Au nanoclusters, since these grafted groups are not stable at T > 1000 K. The annealed nanocluster configurations are depicted in Figure 1

| # | N(Au) | N(X-H) | N(e) | Annealing regime | | |
|---|---|---|---|---|---|---|
| | | | | Starting T, K | Final T, K | Time, ps |
| 1 | 10 | 0 | 790 | 4000 | 0 | 5×100+200 |
| 2 | 18 | 0 | 1422 | 4000 | 0 | 5×100+200 |
| 3 | 38 | 0 | 3002 | 4000 | 0 | 5×100+200 |
| 4 | 20 | 4 OH | 1616 | 500 | 0 | 100+200 |
| 5 | 20 | 4 SH | 1648 | 800 | 0 | 100+200 |
| 6 | 20 | 4 SeH | 1720 | 500 | 0 | 100+200 |

**Methodology**

I implement PM7-MD simulations[14] and annealing simulations to characterize electronic ground-state geometries of the pristine ultrasmall gold nanoclusters. In addition, I consider gold nanoclusters, which are functionalized using a few model chalcogen containing groups, X-H (X = O, S, Se). The provided analysis is based on the molecular dynamics trajectories, which were recorded using an electronic structure description of the nuclear-electronic system. The applied simulation scheme obeys the Born-Oppenheimer approximation, which is widely applied in numerical simulations within the fields of theoretical chemistry and molecular physics. The wave function of all given molecular configurations was optimized repeatedly using the self-consistent field methods (SCF, convergence criterion of $10^{-6}$ Ha). The immediate force acting on each atomic nucleus was derived from the optimized wave function at each time-step. The nuclear trajectory was propagated fully classically via the velocity Verlet integration scheme. Both velocity and position were calculated at the same value of the time variable.

The wave function at each nuclear time-step was optimized using the PM7 ("parametrized model seven") semiempirical Hamiltonian.[20] PM7 is an electronic structure method, which is based on the Hartree-Fock (HF) formalism. It is, therefore, more theoretically fundamental and robust for applications, as compared to empirical force field methods. PM7 was responsible for the electronic part of the calculation to obtain molecular orbitals, heat of formation, and its derivative with respect to molecular geometry. However, as compared to the non-parameterized HF method, PM7 makes a set of approximations and adopts certain parameters in its equations from experimental data to speed up calculations and facilitate wave function convergence. Except serving for faster performance of SCF, empirical parameters are also used to include electron correlation effects. Electron correlation effects are principally omitted in the HF method. PM7 is currently the latest development in the family of semiempirical approaches.[20] It offers a high accuracy of optimized geometries, thermochemistry, band gaps, and electronic spectra. PM7 favorably differs from ab initio electronic structure methods by computational cost.[20] The performance difference comes largely from faster SCF convergence thanks to pre-parameterized integrals. PM7 routinely employs two experimentally determined constants per atom: atomic weight and heat of atomization. Electrostatic repulsion and exchange stabilization are explicitly taken into account. All applicable HF integrals are evaluated by approximate means due to computational efficiency reasons. The set of basis functions consists of one *s* orbital, three *p* orbitals, and five *d* orbitals per each atom. The overlap integrals in the secular equation are ignored.[20]

The nuclear equations-of-motion were integrated with a 0.5 fs time-step for all annealing simulations and with a 1.0 fs time-step for equilibrium-state simulations (at 310 K). The value of 310 K was chosen as a human body temperature in the context of life science applications of gold nanoparticles/nanoclusters. The annealing temperatures (Table 1) were chosen to (1) permit total energy conservation; (2) permit chemical stability of the initial structure. As it was indicated by preliminary simulations, grafted chalcogen containing groups are not stable at the same high temperatures as pristine nanoclusters are, ca. 4000 K. The sulfhydryl groups remain stable at over 1000 K during a few hundreds of picoseconds. In turn, hydroxyl and selenothiol groups leave Au nanocluster extremely fast, as soon as temperature approaches 800 K. This analysis was performed for assessment only and does not pretend to provide exact information on the chemical stability of these derivatives. The selected temperatures (see Table 1) were maintained by the Berendsen thermostat with a relaxation constant of 50 fs.[21]

The latest available revision of MOPAC2012 (obtained from Dr. J.J.P. Stewart) was used to obtain wave functions and forces. The Atomistic Simulation Environment (ASE) set of scripts[22] was used as a starting point to interface electronic structure stage of computations with temperature-coupled velocity Verlet trajectory propagation.

**Results and Discussion**

Five parallel annealing simulations for every system (Table 1) were performed. The configuration, which exhibited the lowest potential energy at the end (Figure 1), was used for further analysis. Other configurations were disregarded. Therefore, all subsequent analysis corresponds to the most stable Au nanocluster. Alternative structures may also exist in nature depending on the method of synthesis and stabilization. Figure 2 summarizes the highest excitation wave lengths for the pristine Au nanoclusters. As should be expected, smaller clusters feature larger band gaps, and therefore, lower HOMO-LUMO excitation wave lengths. Remarkably, $Au_{10}$ and $Au_{18}$ exhibit nearly the same wave length, irrespective of the drastic differences in their shapes. Indeed, $Au_{10}$ is almost planar, while $Au_{18}$ is closer to sphere. Compare, the experimentally determined value for the SPR peak in the case of ca. 50 nm Au nanoparticles is positioned at 520 nm.[13] It is in qualitative concordance with our results. Larger Au nanoclusters must be considered to provide more quantitative extrapolation to 50 nm. This is outside the scope of the present report. In principle, the PM7-MD technique allows consideration of significantly larger Au nanostructures and to sample their thermal motion over reasonable time scales.

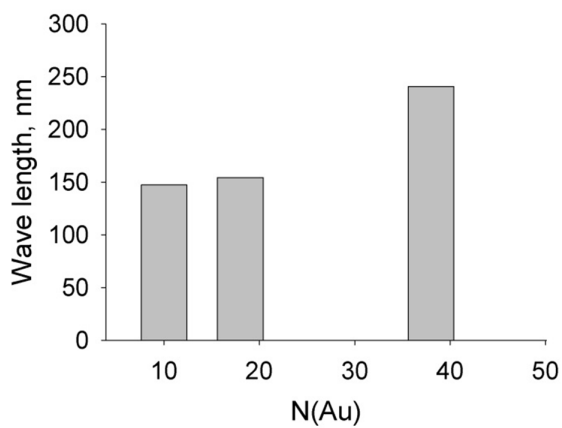

Figure 2. The highest excitation wave length as a function of the number of gold atoms in the pristine gold nanocluster.

Both pristine Au nanoclusters and chalcogen-functionalized nanoclusters were heated up to 310 K. Radial distribution functions (RDFs, Figures 3-4) were calculated using 200 ps long trajectories. The difference between shapes of the $Au_{10}$ and $Au_{18}$ nanoclusters makes it interesting to compare heights and positions of peaks on the corresponding RDFs. The largest and thinnest Au-Au peak occurs in $Au_{10}$, indicating, among other features, its high stability at human body temperature. Therefore, the predicted structure must be considered reliable. Although three repeating Au-Au peaks in $Au_{18}$ and $Au_{38}$ are observed (Figure 3), these structures are not identical to the solid metal structures, because the heights of the peaks decrease drastically. That is, a long-range structure is not well pronounced. Another interesting feature of the RDFs is that the second peak in all nanoclusters is cleaved (at 5.3 Å and 6.0 Å). This constitutes a difference from atom arrangement in the fcc unit cell, which is observed in solid gold.

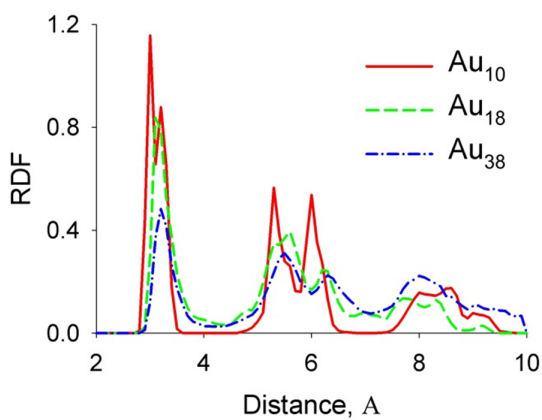

Figure 3. Radial distribution functions, RDFs, computed between gold atoms in the pristine gold nanoclusters. See legend for system designation. RDF normalization is performed in such a way that the integral of each RDF equals to unity.

The hydroxyl and thiol groups provide better coverage (Figure 4). It is indicated by the sharp first peaks at 2.3 Å (oxygen) and 2.5 Å (sulfur). These interatomic distances are significantly smaller than the sum of the van der Waals radii of Au (1.66 Å) and chalcogen elements (1.55 Å for oxygen and 1.80 Å for sulfur). Therefore, covalent binding occurs at the surface of the Au nanocluster, providing a stable functionalization. In turn, the Au-Se first peak is positioned at 3.4 Å, while the van der Waals radius of selenium amounts to 1.90 Å. The Au-Se covalent bond does not form, which can be also due to an extremely high curvature of the $Au_{20}$ surface. The attraction between Au and Se is mainly electrostatic, but the Au-Se ordering is well pronounced (three RDF peaks exhibiting nearly the same height). Nevertheless, selenothiol groups constitute a way less efficient defense for Au nanoparticles, although the resulting structures are still stable in vacuum at 310 K.

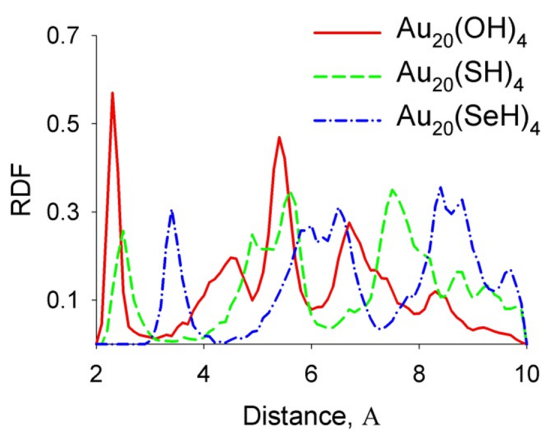

Figure 4. Radial distribution functions, RDFs, computed between gold atoms and chalcogen (O, S, Se) atoms in the chalcogen-functionalized gold nanoclusters. See legend for system designation. RDF normalization is performed in such a way that the integral of each RDF equals to unity.

The smallest simulated nanocluster, $Au_{10}$, represents the most interesting case due to its clearly non-spherical shape (Figure 1), which remains unaltered even during high-temperature annealing. Figure 5 provides an analysis of the electrostatic potential (ESP) at the surface of this cluster (all Au atoms belong to the surface in the present case). The potential was approximated by a set of point charges, which interact to one another via the non-modified Coulomb law. Understanding of ESP is important for prediction of chemical reactivity of the nanostructure as a whole. Large point charges on an arbitrary site may indicate an increased chemical potential. Therefore, a high reactivity must be expected at such a site. Proper functionalization can, to certain extent, solve the reactivity/instability problem.

In spite of the fact that all building blocks forming the cluster (gold atoms) are identical, non-negligible ESP charges are present on almost every Au atom (Figure 5). Two Au atoms possessing equal negative charges (-0.13e) and two Au atoms possessing equal positive charges (+0.09e) occupy symmetric positions (Figure 1). Other Au atoms exhibit smaller charges, both positive and negative ones. Although they are not negligible, the ESP charges at the surface of $Au_{10}$ are not drastically large to point to instability of this ultrasmall nanostructure.

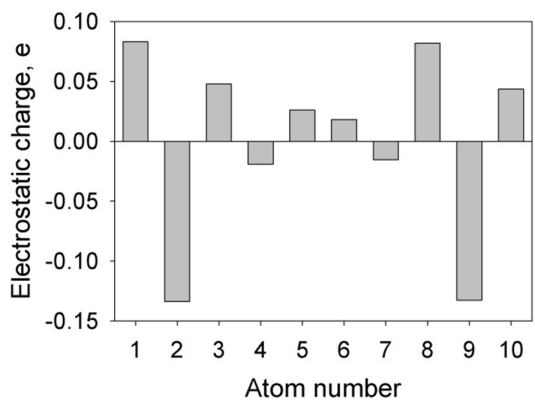

Figure 5. Point Coulombic charges on every gold atom belonging to the $Au_{10}$ nanocluster. The charges were fitted from the electrostatic potential computed at the omega B97XD/LANL2DZ level of theory.[23] The radii of the gold atoms during the fitting procedure were assigned based on the universal force field. Gold atom numbers (x-axis) are assigned arbitrarily.

Figure 6 completely confirms the hypothesis regarding the preferentially electrostatic interaction between selenium atoms and gold atoms. On the contrary, oxygen and sulfur, which create covalent bonds with the Au atoms, exhibit the charges of -0.11e and -0.08e, respectively. By parity of reasoning with low molecular compounds, these covalent bonds can be regarded as low polar ones. Compare, the same level of theory (omega B97XD/LANL2DZ)[23] applied to potassium chloride provides ±0.74e (Hirshfeld charge scheme), which can be regarded as an example of a strong ionic bond with an admixture of polar covalence.

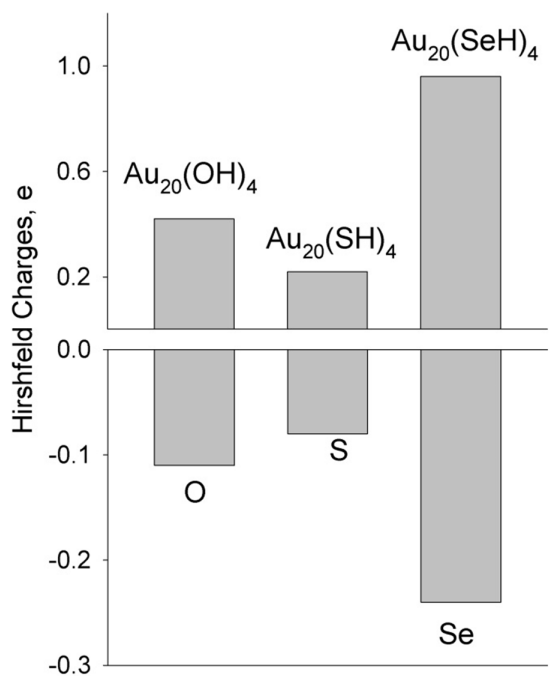

Figure 6. (Top) Deficient electron density in the gold core of the nanocluster computed at the omega B97XD/LANL2DZ level of theory.[23] (Bottom) Excess electron density on each chalcogen atom. The density in both cases was evaluated as a sum of all applicable atom-centered Hirshfeld charges using their classical definition.

**Conclusions**

This work introduces an alternative strategy to obtain trustworthy structures of gold nanoparticles (nanoclusters) employing PM7-MD and annealing simulations. The simulations of $Au_n$ are launched at 4 000 K, which roughly corresponds to the liquid state of gold. Using the Berendsen thermostat, temperature was gradually withdrawn from the system until no kinetic energy left. Five parallel annealing simulations were conducted. Only the final structure with the lowest potential energy was selected, as it is closest to the global energy minimum. Similar procedure was implemented for the functionalized Au nanoclusters (see Table 1 for details) containing –OH, -SH, and –SeH groups. Simulated annealing is the most physically transparent technique to search for stable molecular configurations. Furthermore, it is the only method in the family of geometry optimization algorithms, which has an established analogue in real chemical technology.

The resulting nanostructures were gradually heated up to 310 K and radial distribution functions were used to describe their structure. These results are important for the interpretation of experimental X-ray structures of pristine and functionalized gold nanoparticles. The electron population analysis was carried out using the B97XD/LANL2DZ level of theory,[23] since PM7 does not define electrostatic potential for the elements with d-electrons (gold atoms).[20]

**Acknowledgment**

I thank University of Rochester, New York, United States and personally Drs. Oleg V. Prezhdo and Eric Lobenstine for providing me a courtesy library access outside working hours. MEMPHYS is the Danish National Center of Excellence for Biomembrane Physics. The Center is supported by the Danish National Research Foundation.